\documentclass{article}

\usepackage{PRIMEarxiv}

\usepackage[utf8]{inputenc} 
\usepackage[T1]{fontenc}    
\usepackage{hyperref}       
\usepackage{url}            
\usepackage{booktabs}       
\usepackage{amsfonts}       
\usepackage{nicefrac}       
\usepackage{microtype}      
\usepackage{lipsum}
\usepackage{fancyhdr}       
\usepackage{graphicx}       
\graphicspath{{media/}}     

\usepackage{amsmath}
\usepackage{amssymb}
\usepackage{mathtools}
\usepackage{amsthm}
\usepackage{color}

\usepackage{titletoc}
\usepackage{hyperref,cleveref}
\usepackage{algorithm}
\usepackage{algorithmicx}
\usepackage{algpseudocode}

\title{RL-CFR: Improving Action Abstraction for Imperfect Information Extensive-Form Games with Reinforcement Learning
}

\author{
  Boning Li\\
  IIIS, Tsinghua University \\
  \texttt{li-bn22@mails.tsinghua.edu.cn} \\
  \And
  Zhixuan Fang\\
  IIIS, Tsinghua University \\
  \texttt{fangzhixuan@tsinghua.edu.cn} \\
   \And
  Longbo Huang \thanks{Corresponding author.}\\
  IIIS, Tsinghua University \\
  \texttt{longbohuang@tsinghua.edu.cn} \\
}

\begin{document}
\maketitle

\begin{abstract}
Effective action abstraction is crucial in tackling challenges associated with large action spaces in Imperfect Information Extensive-Form Games (IIEFGs). However, due to the vast state space and computational complexity in IIEFGs, existing methods often rely on fixed abstractions, resulting in sub-optimal performance. In response, we introduce RL-CFR, a novel reinforcement learning (RL) approach for dynamic action abstraction. RL-CFR builds upon our innovative Markov Decision Process (MDP) formulation, with states corresponding to public information and actions represented as feature vectors indicating specific action abstractions. The reward is defined as the expected payoff difference between the selected and default action abstractions. RL-CFR constructs a game tree with RL-guided action abstractions and utilizes counterfactual regret minimization (CFR) for strategy derivation. Impressively, it can be trained from scratch, achieving higher expected payoff without increased CFR solving time. In experiments on Heads-up No-limit Texas Hold'em, RL-CFR outperforms ReBeL's replication and Slumbot, demonstrating significant win-rate margins of $64\pm 11$ and $84\pm 17$ mbb/hand, respectively.
\end{abstract}


\section{Introduction}

The Imperfect Information Extensive-Form Game (IIEFG) model \cite{DBLP:journals/corr/abs-2105-11398} provides a comprehensive framework for analyzing multi-player turn-taking games represented in tree structures. This model encompasses diverse games such as Poker \cite{DBLP:journals/corr/MoravcikSBLMBDW17}, MahJong \cite{DBLP:journals/corr/abs-2307-12087}, and Scotland Yard \cite{2021Player}. Resolving IIEFGs involves determining Nash equilibrium \cite{inbook}, particularly in scenarios featuring two-person zero-sum conditions. In recent years, Counterfactual Regret Minimization (CFR) \cite{DBLP:conf/nips/ZinkevichJBP07} and its variants \cite{DBLP:conf/nips/LanctotWZB09,DBLP:journals/corr/Tammelin14,DBLP:conf/icml/BrownLGS19,DBLP:conf/aaai/BrownS19} have been the predominant methods for addressing large IIEFGs, yielding low-exploitability mixed strategies.

However, numerous IIEFGs exhibit a vast array of actions, leading to an exponential growth in the size of the game tree with the increasing number of actions \cite{DBLP:conf/ijcai/SchnizleinBS09}. This poses a significant computational challenge when applying CFR-based solutions directly. To address this, action abstraction, involving the selection of a limited number of actions from the available set \cite{DBLP:phd/ethos/Aceto91}, has been widely employed to substantially reduce the size of the game tree, facilitating more efficient CFR solving. 
    
Nevertheless, in the realm of IIEFGs,  existing results predominantly focus on fixed action abstractions  \cite{DBLP:journals/corr/MoravcikSBLMBDW17,DBLP:conf/nips/BrownBLG20}. The adoption of fixed action abstractions unavoidably leads to sub-optimality. While methods for dynamic action abstraction exist \cite{DBLP:conf/aaai/HawkinHS11, DBLP:conf/aaai/HawkinHS12,DBLP:conf/aaai/BrownS14}, these approaches often suffer from poor convergence and limited applicability \cite{Brownthesis}. Consequently, identifying strategies capable of achieving dynamic action abstractions with manageable computational complexity remains an outstanding challenge.
    
Reinforcement Learning \cite{DBLP:phd/ethos/Humphreys97, 1998Reinforcement} (RL) has emerged as a revolutionary method in various games such as Go \cite{DBLP:journals/nature/SilverHMGSDSAPL16}, StarCraft II \cite{DBLP:conf/aiide/LeeTZXDA18}, and Dota 2 \cite{DBLP:journals/corr/abs-1912-06680}. However, its application to IIEFGs presents distinctive challenges, primarily arising from two critical features. Firstly, the optimal strategy for an IIEFG typically involves a mixed strategy on its support \cite{DBLP:journals/ijgt/Neyman08}. RL algorithms, however, are primarily designed for learning deterministic policies \cite{DBLP:journals/corr/LillicrapHPHETS15}, making them less suitable for directly handling the probabilistic nature of mixed strategies in IIEFGs. Secondly, the value of an information set in IIEFGs may depend on the chosen strategy \cite{DBLP:conf/nips/BrownBLG20}. However, in RL, the value function is generally assumed to be independent of the agent's policy, posing a challenge in modeling scenarios where the value of an information set is contingent on the specific strategy being employed. Addressing these challenges requires specialized adaptations of RL techniques or exploration of alternative approaches tailored to the unique characteristics of IIEFGs.

\begin{figure}[!htp]
    \begin{minipage}{.64\textwidth}
  \begin{center}
  \includegraphics[width=1\textwidth, trim=100 420 100 30]{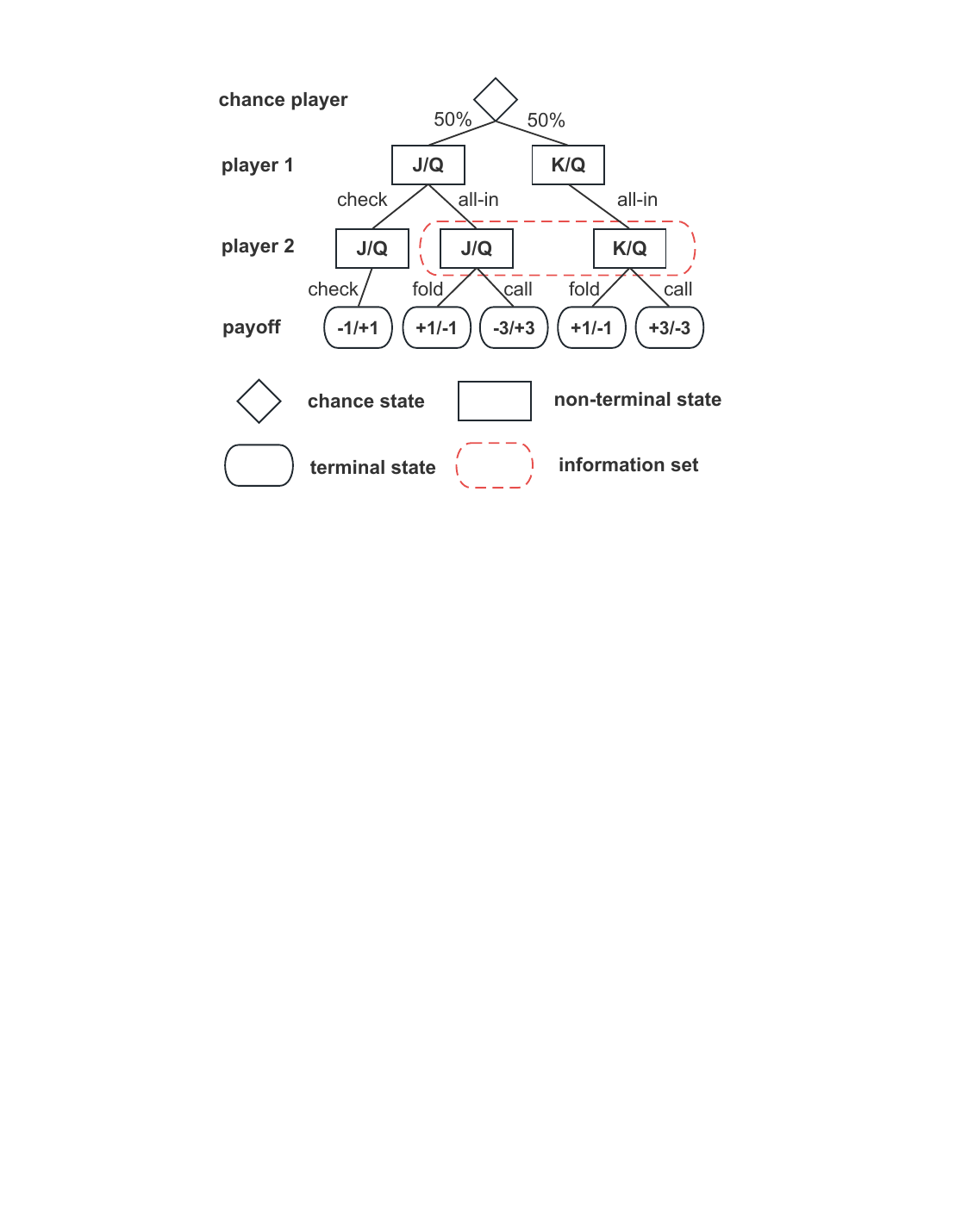}
  \end{center}
  \end{minipage}
  \begin{minipage}{.37\textwidth}
  \caption{
  The game starts with a \emph{chance state} where Player $1$ faces an equal probability of receiving either $J$ or $K$, and Player $2$ is consistently dealt $Q$. If Player $1$ is dealt $K$, an automatic all-in is initiated.  When Player $1$ holds $J$, a pivotal decision arises between opting for a cautious check or committing to an all-in strategy. Subsequently, if Player $1$ opts for an all-in move, Player $2$ is confronted with the dilemma of whether to fold or call. Importantly, Player $2$ remains uninformed about the specific cards held by Player $1$, resulting in an information set that encompasses two states. Upon reaching the terminal state, payoffs are assigned to both players in accordance with a predefined assignment rule. 
  }
  \end{minipage}
  \label{p1}
  \end{figure}

To further clarify these challenges and the motivation for employing RL, consider the simplified poker example \cite{Neilthesis} depicted in Figure \ref{p1}. In this scenario, Player $1$ has an equal chance of being dealt $J$ or $K$, while Player $2$ is always dealt $Q$. Both players contribute $1$ chip to the pot, resulting in a total of $2$ chips, and each player has $2$ chips remaining. Player $1$ acts first. The Nash equilibrium strategy for Player $1$ involves going all-in with $K$ and committing $50\%$ of $J$, while checking the other $50\%$ of $J$. If Player $1$ declares all-in, the Nash equilibrium strategy for Player $2$ is to call and fold with equal probabilities, ensuring that, regardless of Player 1's strategy, Player 2's expected payoff is not reduced. In this equilibrium, Player $1$'s $K$ expects to win $2$ chips, $J$ expects to lose $1$ chip, and Player $2$ expects to lose $0.5$ chips. In contrast, if Player $1$ goes all-in with $100\%$ probability, and Player $2$'s best response strategy is to call with $100\%$ probability, Player $1$'s $K$ expects to win $3$ chips, $J$ expects to lose $3$ chips, and player $2$ expects to win $0$ chips. This example vividly demonstrates the intricate nature of strategies in IIEFGs, where the strategy is likely a mixed strategy, and the chosen strategy impacts the expected values for all players involved.

  \vspace{-0.15in}

To surmount the aforementioned challenges and harness the power of RL in sequential decision making, we introduce RL-CFR, a two-phase framework that ingeniously integrates Deep Reinforcement Learning (DRL) with CFR. In the initial phase, we formulate a novel Markov Decision Process (MDP) \cite{DBLP:books/sp/12/OtterloW12} to identify the action abstraction with the highest expected payoff. Within this MDP, the state encapsulates the \emph{public information} of the game, each control action is represented as a feature vector denoting a specific action abstraction, and action rewards are determined as the payoff differences calculated by CFR between the selected action abstractions and a default fixed action abstraction. 
    
Expanding upon this MDP, we construct a game tree based on the action abstraction chosen by the actor-critic DRL method \cite{DBLP:conf/nips/KondaT99}, ultimately solving the strategy for the selected action abstraction through CFR. RL-CFR offers a principled approach to harness the strengths of both RL and CFR, adeptly addressing challenges related to mixed-strategy and probability-dependent reward scenarios. Importantly, for every state in IIEFG, RL-CFR enhances the expected payoff by selecting a superior dynamic action abstraction compared to a fixed action abstraction without an increase in the CFR solving time or a decrease in convergence. Furthermore, \emph{RL-CFR is capable of being trained from scratch} with only the rules of the IIEFG. This capability is crucial as it allows the algorithm to dynamically adapt to the unique dynamics and intricacies inherent in IIEFGs, mitigating the risk of biases introduced by existing knowledge or inherent biases in the initial model. 
    
To demonstrate the effectiveness of RL-CFR in addressing large IIEFGs, we conducted an evaluation using the challenging Heads-up No-limit Texas Hold'em (HUNL) poker game.\footnote{Heads-up No-limit Texas Hold'em is a two-player variant of Texas Hold'em and serves as a crucial version for exploring mixed strategy two-player zero-sum IIEFGs \cite{DBLP:journals/aim/BardHRZ13}. It is chosen for its intricate nature \cite{DBLP:journals/ai/RubinW11} and an exceedingly vast decision space \cite{DBLP:journals/corr/abs-1302-7008}.} Our results show that RL-CFR outperforms the fixed action abstraction-based HUNL agent ReBeL's replication \cite{DBLP:conf/nips/BrownBLG20} with a win-rate margin of $64$ mbb/hand in a test of over $600,000$ hands. Furthermore, RL-CFR surpasses the well-known HUNL agent Slumbot \cite{2013Slumbot} by a substantial win-rate margin of $84$ mbb/hand in a test of over $250,000$ hands. These substantial win-rate margins underscore the effectiveness of our innovative RL-CFR solution in tackling the challenges posed by large IIEFGs.

The main contributions of our work are  as follows.

    \begin{itemize}

\item \emph{Innovative MDP Formulation for IIEFGs}: We present a pioneering Markov Decision Process (MDP) formulation designed specifically for Imperfect Information Extensive-Form Games (IIEFGs). This formulation meticulously defines states based on public information, represents actions as feature vectors denoting action abstractions, and establishes rewards as value differences between selected action abstractions and default fixed action abstractions. The dynamic adjustment of action abstractions at different states is facilitated by our MDP formulation, enhancing adaptability.

\item \emph{RL-CFR Framework Integration}: Building upon our novel MDP, we introduce the RL-CFR framework—a novel fusion of Deep Reinforcement Learning (DRL) with Counterfactual Regret Minimization (CFR). This framework achieves a harmonious balance between computation and optimism, and can be trained from scratch with only the rules of the IIEFG. RL-CFR adeptly addresses the challenges posed by the large decision space and computational complexity inherent in IIEFGs, offering a customizable tradeoff mechanism between CFR-related computational complexity and RL-driven performance improvement.

\item \emph{Evaluation on HUNL Poker Game}: We evaluate the performance of RL-CFR on the widely recognized Heads-up No-limit Texas Hold'em (HUNL) poker game. Our results showcase the superiority of RL-CFR by achieving significant win-rate advantages over ReBeL's replication, one of the leading fixed action abstraction-based HUNL algorithms, and Slumbot, a robust publicly available HUNL AI for online comparisons. Specifically, RL-CFR outperforms ReBeL's replication and Slumbot by substantial win-rate margins of $64\pm 11$ and $84\pm 17$ mbb/hand, respectively. 
\end{itemize}

\section{Related Work on Extensive-Form Games}

  \textbf{Methods for solving IIEFGs.}
CFR-based algorithms, such as those proposed by \cite{DBLP:journals/corr/abs-1303-4441,DBLP:journals/corr/Tammelin14,DBLP:conf/aaai/BrownS19,DBLP:conf/icml/BrownLGS19}, are commonly employed for solving large IIEFGs. The choice of CFR is attributed to its advantageous linear bounding of regret concerning the game size, with a more detailed description available in Appendix \ref{cfr}. Additionally, alternative methods such as Hedge \cite{DBLP:books/daglib/0016248} and the excessive gap technique \cite{DBLP:journals/mor/HodaGPS10} theoretically demonstrate faster convergence than CFR.

\textbf{Action Abstraction in IIEFGs.}
The action abstraction technique expedites strategy computation in IIEFGs, providing solutions with theoretical bounds \cite{DBLP:conf/sigecom/KroerS14, DBLP:conf/nips/KroerS18, DBLP:conf/atal/KroerS15}. In IIEFGs with a multitude of actions, such as poker games, the impact of action abstraction on strategy quality can be surprisingly significant \cite{DBLP:conf/atal/WaughSBS09, 2007The}. Parametric methods proposed by \cite{DBLP:conf/aaai/HawkinHS11,DBLP:conf/aaai/HawkinHS12} aim to find optimal action abstraction for early states of IIEFGs, while an iterative algorithm \cite{DBLP:conf/aaai/BrownS14} has been introduced to adjust action abstraction during iteration. However, it is worth noting that these methods, altering action abstraction during CFR iteration, tend to converge more slowly compared to fixed action abstraction methods \cite{Brownthesis}.

\textbf{Reinforcement Learning (RL) Approaches for IIEFGs.} Various methodologies inspired by RL have emerged to solve IIEFGs. Noteworthy contributions include regression counterfactual regret minimization \cite{DBLP:conf/aaai/WaughMBB15,DBLP:conf/atal/DOrazioMWB20}, neural fictitious self-play \cite{DBLP:journals/corr/HeinrichS16}, and ReBeL \cite{DBLP:conf/nips/BrownBLG20}. Furthermore, \cite{DBLP:conf/icml/PerolatMLOROBAB21} introduced a regularization-based reward adaptation technique, ensuring robust convergence guarantees when addressing two-player zero-sum IIEFGs. In addition, \cite{DBLP:conf/iclr/LiuOYZ23} delved into RL regularization techniques for IIEFGs, proposing a regularization-based payoff function. To tackle the challenge of inaccurate state value estimation in large IIEFGs, \cite{DBLP:conf/aaai/MengGTA023} presented an efficient deep reinforcement learning method. Additionally, \cite{anonymous2024dynamic} enhanced the DCFR algorithm \cite{DBLP:conf/aaai/BrownS19} by replacing the fixed discount parameter with a dynamic discount parameter using RL selection, resulting in improved outcomes. These approaches collectively underscore the diverse applications of RL methodologies in addressing various challenges within the domain of IIEFGs.

\section{Background and Notation}

    \label{back}

    \textbf{Imperfect Information Extensive-Form Games.} An Imperfect Information Extensive-Form Game (IIEFG) \cite{DBLP:journals/corr/abs-2105-11398} models the sequential interaction of one or more players. Let $\mathcal N=\{1,\cdots,N\}$ be the set of players, and $H$ be the set of states (histories). A state $h\in \mathcal{H}$ is defined by all historical actions from the initial game state $\emptyset$. The state $h\cdot a\in \mathcal H$ is a child of the state $h$, and $h\sqsubseteq h'$ implies that $h$ is an ancestor of $h'$. $\mathcal Z$ is the set of terminal states. For each non-terminal state $h$, $\mathcal A(h)$ is the set of available actions, $\mathcal {AA}(h)\subseteq \mathcal A(h)$ is an \emph{action abstraction} for $\mathcal A(h)$, and $\mathcal P(h)\in\mathcal N\cup\{c\}$ determines the acting player, where $c$ denotes the ``chance player," representing random events players in $\mathcal N$ cannot control. $\sigma_c(h,a)$ is the probability that chance player will act $a\in\mathcal A(h)$ at state $h$. $\mathcal H_p$ is the set of all states $h$ satisfy $\mathcal P(h)=p$. For every $z\in\mathcal Z$, $u=(u_p)_{p\in\mathcal N}$ gives the payoff for each players.

    The information-partition $\mathcal I=(\mathcal I_p)_{p\in\mathcal N}$ describes the imperfect information of the IIEFG, where $\mathcal I_p$ is a partition of $\mathcal H_p$ for each player $p$. A set $I\in\mathcal I_p$ is called an information set, and all states in $I$ are indistinguishable for player $p$. We denote $I(h)$ as the unique information set that contains $h$. A behaviour strategy $\sigma_p\in\Sigma_p$ is a function where $\sigma_p(I,a)\in\mathbb R$ determines the probability distribution over available actions $a\in\mathcal A(I)$ for every information set $I\in\mathcal I_p$. We denote $\sigma(I,a)=\sigma_{\mathcal P(I)}(I,a)$. $\sigma=(\sigma_p)_{p\in\mathcal N}$ is a strategy profile.
    $\pi^{\sigma}(h)$ is the probability of reaching state $h$ if players follow $\sigma$, calculated as $\pi^{\sigma}(h)=\prod_{h'\cdot a\sqsubseteq h}\sigma(h',a)$. The counterfactual value (CFV) \cite{DBLP:conf/nips/ZinkevichJBP07} of an information set $I\in\mathcal I_p$ is the expected utility for player $p$ given that $I$ has been reached, calculated as $v_p^{\sigma}(I)=\sum_{h\in I}(\pi_{-p}^{\sigma}(h)\sum_{z\in\mathcal Z,h\sqsubseteq z}(\pi^{\sigma}(z|h)u_p(z)))$.

    \textbf{Public Belief State.} A Public Belief State (PBS) is an extended notion of history for IIEFGs based on the common knowledge belief distribution over histories \cite{DBLP:conf/aaai/BurchJB14,DBLP:conf/atal/SustrKL19}. Specifically, we define player $p$'s observation-action history (infostate) \cite{DBLP:conf/aaai/BurchJB14} as $O_p=(I_1,a_1,I_2,a_2,\cdots)$, which includes the information sets visited and actions taken by player $p$. The unique infostate corresponding to a state $h\in\mathcal H_p$ for player $p$ is $O_p(h)$. The set of states that correspond to $O_p$ is denoted $\mathcal H(O_p)$. We use $\sim$ to denote states indistinguishable by some player, i.e.,  $g\sim h$ means $\bigvee_{i=1}^N O_i(g)=O_i(h)$ ($\bigvee$ is the OR operation on all expressions).
    
        A public partition is any partition $\mathcal {PS}$ of $\mathcal H\backslash\mathcal Z$ whose elements are closed under $\sim$ and form a tree \cite{DBLP:conf/ijcai/JohansonWBZ11}. An element $PS\in\mathcal {PS}$ is called a public state that includes the public information that each player knows. The unique public state for a state $h$ and an infostate $O_p$ are denoted by $PS(h)$ and $PS(O_p)$, respectively. The set of states that match the public information of $PS$ is denoted as $\mathcal H(PS)$.

    In general, a PBS $\beta$ is described by the joint probability distribution of the possible infostates of the players \cite{2013Decentralized, DBLP:conf/ijcai/Oliehoek13, DBLP:journals/jair/DibangoyeABC16}. 
    Formally, given a public state $PS$, $\mathcal O_p(PS)$ is the set of infostates that player $p$ may be in, and $\triangle \mathcal O_p(PS)$ is a probability distribution over the elements of $\mathcal O_p(PS)$. Then, PBS $\beta=(\triangle \mathcal O_1(PS),\cdots,\triangle \mathcal O_N(PS))$. The public state of PBS $\beta$ is denoted as $PS(\beta)$. The acting player at PBS $\beta$ is denoted $\mathcal P(\beta)$. The set of available actions for acting player at PBS $\beta$ is denoted $\mathcal A(\beta)$, and the action abstraction at PBS $\beta$ is denoted $\mathcal {AA}(\beta)$.

    A subgame can be rooted at a PBS because PBS is a state of the perfect-information belief-representation game with well-defined values \cite{DBLP:journals/corr/abs-1906-06412,DBLP:conf/nips/BrownBLG20}. At the beginning of a subgame, a history is sampled from the probability distribution of the PBS, and then the game plays as if it is the original game. The value for player $p$ of PBS $\beta$ (PBS value) when all players play according to $\sigma$ is defined as $v_p^{\sigma}(\beta)=\sum_{h\in\mathcal H(PS(\beta))}(\pi^{\sigma}(h|\beta)v_p^{\sigma}(h))$. The value for an infostate $O_p\in\beta$ when all players play according to $\sigma$ is defined as $v_p^{\sigma}(O_p|\beta)=\sum_{h\in\mathcal H(O_p)}(\pi^{\sigma}(h|O_p,\beta_{-p})v_p^{\sigma}(h))$ where $\pi^{\sigma}(h|O_p,\beta_{-p})$ is the probability of reaching state $h$ according to $\sigma$, assuming $O_p$ is reached and the probability distribution over infostates for players other than $p$ is $\beta_{-p}$. 

\section{A Novel MDP Formulation for IIEFGs}

    \label{mdp}

    In this section, we present our novel MDP formulation tailored for IIEFGs. It is important to emphasize that our formulation serves as an abstract MDP model, strategically designed to determine the action abstraction of IIEFGs. This action abstraction, once determined, becomes the basis for executing a CFR algorithm, allowing us to solve for the mixed strategy. In this MDP, states correspond to public information, and actions are represented as feature vectors indicating specific action abstractions. The reward is defined as the expected payoff difference between the selected and default action abstractions. Below, we delve into the specific details of our MDP.

    In general, a Markov Decision Process (MDP) \cite{DBLP:books/sp/12/OtterloW12} comprises the tuple $\langle \mathbf{S},\mathbf{A},P,r,\gamma\rangle$, where $\mathbf{S}$ is the set of states, $\mathbf{A}$ is the set of actions, $r:\mathbf{S}\times \mathbf{A}\mapsto \mathbb R$ is the reward function, $P(\mathbf s'|\mathbf s,\mathbf a)$ is the state transfer function, and $\gamma$ is the discount factor. The primary objective is to find an optimal control policy $\pi^{*}$, which determines $\mathbf a_t=\pi^{*}(\mathbf s_t)$ at each time, aiming to maximize the expected cumulative reward $R=\mathbb{E}\{\sum_{t=0}^{\infty}\gamma^tr(\mathbf s_t,\mathbf a_t)\}$.

    \textbf{New state, action and reward function for IIEFGs.}

    (\textbf{State}) 
    State $\mathbf s=PS(\beta)$, where $PS(\beta)$ is the public state of the current PBS $\beta$. Our design offers two notable advantages compared to using PBS $\beta$ directly as a state: 
    
    1. \emph{Dimension Reduction.} Our state effectively reduces the dimensionality. PBS typically has a large dimension as it needs to record the distribution of infostates. In contrast, the public state has a more moderate dimension since it only captures public information known to all players.\footnote{For example, in HUNL, a public state includes only essential information like previous actions of the players, public cards, chips in the pot, remaining chips and the acting player. In contrast, a public belief state encompasses $1,326$ different private hands for both players, and requires at least $2,652$ dimensions.} 
    
    2. \emph{Stability during CFR iterations.} The public states of the non-root nodes remain fixed during CFR iterations. In the iterative process of CFR, PBS of the non-root nodes may change at each iteration, while the public state remains constant. Since the selection of the action abstraction is based on the state, maintaining a fixed state during CFR iterations is crucial. If the state changes at each iteration, the action abstraction will also change, potentially leading to poor convergence of CFR \cite{Brownthesis}.

     (\textbf{Action}) Action $\mathbf a=(x_1,y_1,\cdots,x_K,y_K)$ is a $2K$-dimensional vector, where $K$ corresponds to the number of actions that can be selected in the chosen action abstraction. Each $x_i,y_i$ have values between $-1$ and $1$. In our MDP, this action $\mathbf{a}$ is utilized to select an action abstraction $\mathcal{AA}_{\text{MDP}}(\beta,\mathbf{a})$ at PBS $\beta$. The subsequent game tree for CFR solving is constructed based on this action abstraction. The specifics are detailed below.

     In an IIEFG, there are actions that are common and are added to the action abstraction regardless of the PBS $\beta$. We denote this set of actions as always-selected action set $\mathcal{AA}_{\text{always}}$, which may include some of the most common actions available. Additionally, we define a \emph{default} fixed action abstraction $\mathcal{AA}_{\text{base}}(\beta)$ at PBS $\beta$. Here, $\mathcal{AA}_{\text{base}}(\beta)\subseteq \mathcal{A}(\beta)$ is a set of actions solely related to PBS $\beta$, and we have $\mathcal{AA}_{\text{always}}\subseteq \mathcal{AA}_{\text{base}}(\beta)$. 
     Typically, $\mathcal{AA}_{\text{base}}(\beta)$ is pre-specified to a set of available actions related to crucial information of PBS $\beta$. 

     The choices for $\mathcal{AA}_{\text{always}}$ and $\mathcal{AA}_{\text{base}}(\beta)$ can be arbitrary in any IIEFG. However, different choices can impact the win-rate and running time, as demonstrated in \cite{DBLP:journals/corr/MoravcikSBLMBDW17}. For examples, in HUNL experiments, when $\mathcal{AA}_{\text{always}}=\{F,C,A\}$ ($F,C,A$ refer to fold, check/call and all-in respectively), \cite{DBLP:journals/corr/MoravcikSBLMBDW17} shows that the action abstraction $\mathcal{AA}_{\text{base}}(\beta)=\mathcal{AA}_{\text{always}}\cup \{0.5\times \text{pot},1\times \text{pot},2\times \text{pot}\}$ ($\times \text{pot}$ means the fraction of the size of the pot being bet) achieves a win-rate of $96$ mbb/hand compared to action abstraction $\mathcal{AA}_{\text{base}}(\beta)=\mathcal{AA}_{\text{always}}\cup \{1\times \text{pot}\}$. 
     
    Formally at PBS $\beta$, the action abstraction chosen by $\mathbf{a}=(x_1,y_1,\cdots,x_K,y_K)$ is 
     \begin{equation}
    \mathcal {AA}_{\text{MDP}}(\beta,\mathbf{a})=\mathcal{AA}_{\text{always}}\cup\mathcal{AA}_{\text{optional}}(\beta,\mathbf{a})
     \end{equation}
     
     Here, the optional action set $\mathcal{AA}_{\text{optional}}(\beta,\mathbf{a})$ is the set of actions generated from PBS $\beta$ and the chosen action vector $\mathbf{a}$. Since the size of the game tree increases exponentially with the number of available actions, we limit the optional action set $\mathcal{AA}_{\text{optional}}(\beta,\mathbf{a})$ to have most $K$ actions. Precisely, $\mathcal{AA}_{\text{optional}}(\beta,\mathbf{a})=\bigcup_{i=1}^K f(x_i,y_i,\beta)$, where $f(x_i,y_i,\beta)$ is a function that generates an available action from all available actions except those in $\mathcal{AA}_{\text{always}}$. Notably, if $f(x_i,y_i,\beta)=\emptyset$, it means that there is no chosen action in this dimension of the action abstraction. Since the set of available actions of IIEFGs with numerous actions tends to be continuous, we define the function $f(x_i,y_i,\beta)$ by correspondingly mapping continuous parameters $x_i$ and $y_i$ to the set of available actions $\mathcal{A}(\beta)$ of PBS $\beta$.

     Below, using HUNL as an example, we describe how to choose $K$ and define $f(x_i,y_i,\beta)$. We set $K=3$, which means we can select up to $3$ raising scales other than all-in. We let $\mathcal{AA}_{\text{always}}=\{F,C,A\}$ and $\mathcal{AA}_{\text{base}}(\beta)=\{F,C,A,0.5\times \text{pot},1\times \text{pot},2\times \text{pot}\}$ (consistent with \cite{DBLP:journals/corr/MoravcikSBLMBDW17}). Based on human experience and inspired by prior studies \cite{DBLP:conf/aaai/HawkinHS11, DBLP:conf/aaai/HawkinHS12}, a reasonable range for a raising scale other than all-in is $[0,5]\times \text{pot}$. Thus, we define the $f(x_i,y_i,\beta)$ function to be 
    \begin{equation}
f(x_i,y_i,\beta)=\begin{cases}
\text{CLIP}(2.5(x_i+1)\times \text{pot}),&y_i\geq 0;\\
\emptyset,&y_i<0.
\end{cases}
\end{equation}
    where $\text{CLIP}$ is a function that corresponds the nearest available raising scale.


     (\textbf{Reward}) The reward function $r(\mathbf s,\mathbf a)$ in our MDP, is defined as the difference in PBS values between the chosen action abstraction $\mathcal{AA}_{\text{MDP}}(\beta,\mathbf a)$ and the default action abstraction $\mathcal{AA}_{\text{base}}(\beta)$.

     For our abstract MDP, each action's reward needs to be obtained by solving two independently depth-limited subgames \cite{DBLP:conf/nips/BrownSA18} using the CFR-based ReBeL algorithm \cite{DBLP:conf/nips/BrownBLG20}, as described in Appendix \ref{app1}. Here is how we compute the reward $r$ based on the PBS $\beta$, the state vector $\mathbf s$, and the action vector $\mathbf a$:

     1. \emph{Build Game Tree.} Construct a game tree rooted at PBS $\beta$ with selected action abstraction $\mathcal{AA}_{\text{MDP}}(\beta,\mathbf a)$. Note that this selected action abstraction is used only for the root.

     2. \emph{Compute Strategy Profile $\sigma_{\text{MDP}}$.} Utilize ReBeL to obtain a strategy profile $\sigma_{\text{MDP}}$ for PBS $\beta$. This profile provides state transfers for all infostates corresponding to non-leaf nodes in the subgame.

     3. \emph{Calculate PBS Value for Chosen Action Abstraction.}  With the calculated strategy profile $\sigma_{\text{MDP}}$, compute the PBS value $v_{\mathcal P(\beta)}^{\sigma_{\text{MDP}}}(\beta)$ for the acting player. This represents the expected payoff calculated for the acting player on PBS $\beta$ (details of the PBS value calculation are in the last paragraph of Section \ref{back}).

     4. \emph{Build Another Game Tree.} Construct another game tree rooted at PBS $\beta$, but this time using the default fixed action abstraction $\mathcal{AA}_{\text{base}}(\beta)$ at the root. For the non-root nodes, both game trees use either the default action abstraction or the MDP-based action abstraction, depending on the training stage. Further details about the choice of action abstraction for non-root nodes can be found in Sections \ref{p5} and \ref{ex}.

     5. \emph{Compute Strategy Profile $\sigma_{\text{base}}$.} Similarly, obtain the strategy profile $\sigma_{\text{base}}$ for this game tree based on ReBeL.

     6. \emph{Calculate PBS Value for Default Action Abstraction.} Using the calculated strategy profile $\sigma_{\text{base}}$, compute the PBS value $v_{\mathcal P(\beta)}^{\sigma_{\text{base}}}(\beta)$ for the acting player.

     7. \emph{Compute Reward.} Finally, compute the reward $r(\mathbf s,\mathbf a) = v_{\mathcal P(\beta)}^{\sigma_{\text{MDP}}}(\beta) - v_{\mathcal P(\beta)}^{\sigma_{\text{base}}}(\beta)$.

      The state transition of the MDP depends on the mixed strategy calculated by the CFR algorithm, as detailed in Section \ref{p5}. The discount factor $\gamma$ is set to $1$ during training.

\section{RL-CFR Framework}
    \label{p5}

    In this section, we introduce the RL-CFR framework, an extension of ReBeL algorithm \cite{DBLP:conf/nips/BrownBLG20}. The ReBeL algorithm is known for its efficiency in solving depth-limited subgames \cite{DBLP:conf/nips/BrownSA18}, allowing it to obtain effective fixed action abstraction strategies for large IIEFGs through a combination of self-play RL and CFR (see Appendix \ref{app1}). In contrast to ReBeL, which employs a fixed action abstraction, RL-CFR takes a novel approach by dynamically selecting its action abstraction through RL, based on our novel MDP. 
    
    As demonstrated in our experiments, this dynamic selection allows for the discovery of superior action abstractions, resulting in significant performance improvements. It is noteworthy that applying the DRL approach to IIEFGs is a highly nontrivial task. The main challenge lies in determining a mixed strategy for all information sets \cite{Neilthesis, Brownthesis}, a computation that is intricate when approached directly through RL methods. Despite this challenge, RL-CFR tackles the complexity by incorporating dynamic action abstraction, showcasing its capability to navigate the intricacies of IIEFGs and deliver enhanced performance outcomes.

\begin{figure}[ht]
  \centering

  \includegraphics[width=0.8\textwidth]{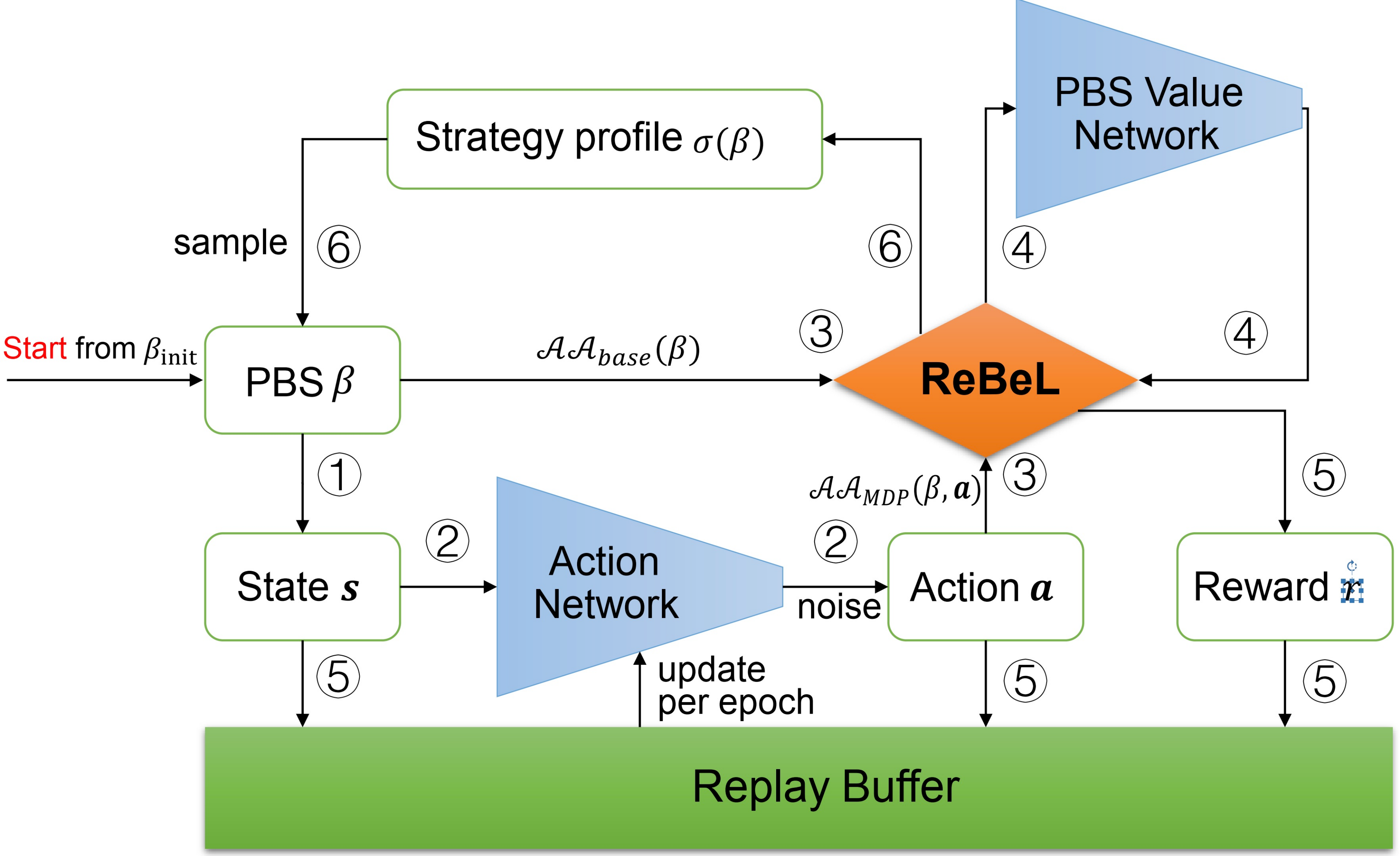}
  \caption{
  Training procedure for the RL-CFR framework. The labels in the figure correspond to the sampling steps for RL-CFR framework. A sampling epoch starts from the initial PBS $\beta_{\text{init}}$.}
  \label{p5:2procedure}
  \end{figure}

The RL-CFR framework constitutes an end-to-end self-training reinforcement learning process, as illustrated in Figure \ref{p5:2procedure}. We now elucidate the sampling steps for RL-CFR framework:\footnote{Here, ``Sampling" refers to the process of selecting and generating instances or experiences that are used for training and updating the RL model.} 

\emph{Step \textcircled{\small{1}} Compressing PBS.} Starting from the initial PBS of the game, each handling of a PBS $\beta$ involves several stages.  If we encounter a chance PBS where the acting player is a chance player, we allow the chance player to act randomly and update the PBS. If we encounter a terminal PBS, the epoch of sampling ends. Moving on to a non-chance and non-terminal PBS $\beta$, we proceed to compress the high-dimensional PBS $\beta$ into a low-dimensional public state $\mathbf{s}$ using the method described in Section \ref{mdp}.

\emph{Step \textcircled{\small{2}} Action Abstraction Selection.} Passing through the action network and adding a Gaussian noise (for increased exploration) yields an action vector $\mathbf{a}$, which is then mapped to a specified action abstraction $\mathcal {AA}_{\text{MDP}}(\beta,\mathbf{a})$. 

\emph{Step \textcircled{\small{3}} Building Depth-Limited Subgames.} Constructing two depth-limited subgames rooted at $\beta$ according to the default action abstraction $\mathcal {AA}_{\text{base}}(\beta)$ and selected action abstraction $\mathcal {AA}_{\text{MDP}}(\beta,\mathbf{a})$, respectively. 

\emph{Step \textcircled{\small{4}} ReBeL Algorithm.} Utilizing the ReBeL algorithm to solve the strategies and values of the two subgames. 

\emph{Step \textcircled{\small{5}} Calculating Reward and Updating RL Data.} Calculating the PBS value difference as a reward $r$, and adding RL data $\{\textbf{s},\textbf{a},r\}$ to the training data (denoted as $Data^{\text{RL}}$) for the action and critic network. 

\emph{Step \textcircled{\small{6}} State Transition.} Randomly choosing a subgame and following the corresponding strategy $\sigma(\beta)$ for state transition to a child PBS $\beta'$ next. Setting $\beta=\beta'$, and repeating Step \textcircled{\small{1}}. 

Algorithm \ref{a1} shows the formal procedure of the sampling process. These steps collectively form the foundation of the RL-CFR sampling process, driving the iterative learning and optimization within the framework.

     \begin{algorithm}[!h]
     \label{a1}
    \caption{RL-CFR framework: Sampling $(s,a,r)$ data}
    \begin{algorithmic}
    
    \Require $\theta_{\alpha},\text{noise},\eta,\epsilon$ \Comment{\textcolor{blue}{$\text{Noise}=0.15,\eta=0.33,\varepsilon=0.25$ during training}}
    
    \State $\beta \leftarrow \beta_{\text{init}}$ 
    
    \State $Data^{\text{RL}}\leftarrow \{\}$

    \While{!IsTerminal$(\beta)$}

        \While{$P(\beta)=c$}
        
            \State $\beta\leftarrow$ TakeChance$(\beta)$ \Comment{\textcolor{blue}{Random chance event}}
        
        \EndWhile
        
        \State $\sigma_{\text{base}}(\beta), v_{\mathcal P(\beta)}^{\sigma_{\text{base}}}(\beta)
        \leftarrow$ReBeL$(\beta, \mathcal {AA}_{\text{base}}(\beta))$ \Comment{\textcolor{blue}{Compute the strategy and value for default action abstraction}}

        \State $s\leftarrow PS(\beta)$ \Comment{\textcolor{blue}{Use public state as the state in MDP}}
        
        \State $\mathbf{a}\leftarrow$ActionNetwork$(s,\theta_{\alpha})+\mathcal N(0,\text{noise})$ 

        \State $\sigma_{\text{MDP}}(\beta), v_{\mathcal P(\beta)}^{\sigma_{\text{MDP}}}(\beta)\leftarrow$ReBeL$(\beta, \mathcal {AA}_{\text{MDP}}(\beta,\mathbf{a}))$ \Comment{\textcolor{blue}{Compute the strategy and value for selected action abstraction}}

        \State $r\leftarrow v_{\mathcal P(\beta)}^{\sigma_{\text{MDP}}}(\beta)-v_{\mathcal P(\beta)}^{\sigma_{\text{base}}}(\beta)$\Comment{\textcolor{blue}{Reward function}}

        \State Add $\{\textbf{s},\textbf{a},r\}$ to $Data^{\text{RL}}$

        \State $c\sim \text{unif}[0,1]$

        \State $d\sim \text{unif}[0,1]$

        \If{$c<\eta$}
        
            \If{$d<\epsilon$}
            
                \State $\text{nextaction}\sim \mathcal {AA}_{\text{base}}(\beta)$
                
            \Else
            
                \State $\text{nextaction}\sim \sigma_{\text{base}}(\beta)$

            \EndIf

            \State $\beta\leftarrow \text{NEXTPBS}(\beta,\sigma_{\text{base}}(\beta),\text{nextaction})$

        \Else
        
            \If{$d<\epsilon$}
            
                \State $\text{nextaction}\sim \mathcal {AA}_{\text{MDP}}(\beta,\mathbf{a})$
                
            \Else
            
                \State $\text{nextaction}\sim \sigma_{\text{MDP}}(\beta)$
                
            \EndIf
            
            \State $\beta\leftarrow \text{NEXTPBS}(\beta,\sigma_{\text{MDP}}(\beta),\text{nextaction})$
            
        \EndIf

    \EndWhile

    \Ensure{$Data^{\text{RL}}$}
    \end{algorithmic}
    \end{algorithm}

        After each epoch, a trajectory $(\mathbf{s}_1, \mathbf{a}_1, r_1, \cdots, \mathbf{s}_t, \mathbf{a}_t, r_t)$ is sampled based on the current action network, where $t$ is the length of the game, contingent on the player actions. Upon collecting RL data $Data^{\text{RL}}=\{(\mathbf{s},\mathbf{a},r)\}$ over several epochs, the Actor-Critic algorithm \cite{DBLP:conf/nips/KondaT99} is employed, along with Mean Squared Error (MSE) Loss to train both the action network and critic network (refer to Section \ref{ex} for network structures). The loss functions are defined as follows: 
     \begin{eqnarray}
     \mathcal L(\theta_c)=\mathbb E_{(\mathbf{s},\mathbf{a},r)\sim Data^{\text{RL}}}[(r^{\theta_c}(\mathbf{s},\mathbf{a})-r)^2],\\ \quad \mathcal L(\theta_a)=\mathbb E_{(\mathbf{s},\mathbf{a},r)\sim Data^{\text{RL}}}[-r^{\theta_c}(\mathbf{s},\mathbf{a}^{\theta_a}(s))], 
     \end{eqnarray}
     where $\theta_c, \theta_a$ represent the parameters of the critic network and action network, respectively. 
     
    During the initial epochs of training, the action network may tend to select suboptimal action abstractions compared to the default fixed action abstraction $\mathcal{AA}_{\text{base}}(\beta)$. To address this, we initiate training by utilizing the default action abstraction for non-root nodes when constructing the depth-limited subgame. As training progresses, the action network becomes more adept at selecting superior action abstractions. Consequently, we transition to choosing the action abstraction for non-root nodes based on the output of the action network when building the depth-limited subgame. Simultaneously, to enhance the accuracy of PBS values, we can retrain the PBS value network according to the action abstraction selected by the action network. The iterative process of updating the PBS value network and the action network can theoretically be repeated for ongoing training, allowing the framework to adapt and improve its decision-making capabilities over time.

\section{Experiment}
\label{ex}

    To demonstrate the effectiveness of our RL-CFR framework in handling large IIEFGs with numerous actions, we conducted experiments on Heads-up No-limit Texas Hold'em (HUNL). Similar to prior studies on large IIEFGs \cite{DBLP:conf/icml/BrownLGS19,DBLP:conf/nips/BrownBLG20, DBLP:journals/corr/abs-2007-10442}, we chose HUNL due to its representative nature and inherent complexity. The rules of HUNL are provided in Appendix \ref{rule}. During evaluation, players start with $200$ big blinds, switching positions every two hands, akin to the annual computer poker competition (ACPC) \cite{DBLP:journals/aim/BardHRZ13}.

    Experiments were executed on a server with 8 NVIDIA PH402 SKU 200 GPUs and an 80-core Intel(R) Xeon(R) Gold 6145 2.00GHz CPU. Neural networks in our implementation, including the PBS value network, action network, and critic network, are Multi-Layer Perceptrons (MLPs) with ReLU activation functions. The networks are trained with the Adam optimizer \cite{DBLP:journals/corr/KingmaB14}. In the CFR iteration to solve a PBS, we use the discounted CFR (DCFR) algorithm \cite{DBLP:conf/aaai/BrownS19}, with the number of iterations $T=250$ during training and evaluation.\footnote{Since HUNL evaluations are generally time-limited and need to be solved within a few seconds, common HUNL AIs typically take between $100$ to $1000$ CFR iterations \cite{DBLP:conf/aaai/BrownGS15a, DBLP:conf/nips/BrownBLG20,DBLP:conf/ijcai/BrownS17,DBLP:journals/corr/MoravcikSBLMBDW17}.} 

    We initiated our experiments by training a PBS value network, comprising approximately 18 million parameters (as detailed in Appendix \ref{app1}). It is worth noting that all PBS value networks used in our experiments, including those employed in RL-CFR, were trained based on the default action abstraction. This deliberate setting aims to emphasize that any performance enhancements achieved by RL-CFR are solely attributed to the action abstraction chosen by the action network.

    The action network and the critic network both have $3$ layers and $2\times 10^4$ parameters, with hidden layers of dimensions $128$ and $96$. The training process has $2\times 10^6$ epochs, each sampling approximately $10$ RL data.\footnote{An RL data instance $(\mathbf{s},\mathbf{a},r)$ comprises a $64$-dimensional state $\mathbf{s}$ (recording public cards, chips, positions and previous actions in HUNL), a $6$-dimensional action $\mathbf{a}$ and a scalar reward $r$. Since the number of rounds in a HUNL game is not deterministic, a single sample to the terminate state generally yields no more than $10$ pieces of RL data.} Random sampling from the entire RL data set was conducted, utilizing a learning rate of $1\times 10^{-5}$ and a batch size of $1,024$ during training. After $5\times 10^5$ epochs, we generated PBS data by constructing the game tree precisely according to the action abstraction provided by the action network. Notably, the training cost of action network and critic network is approximately $30$\% of the training cost of PBS value network.

    We assessed the head-to-head performance of RL-CFR against the replication of ReBeL under the common knowledge setting. In this context, common knowledge entails that each player possesses information about their opponent's strategy and previous actions during the hand , and such setting does not affect the evaluation \cite{DBLP:conf/aaai/BurchSMMB18} and circumvents the nested subgame solving technique \cite{DBLP:conf/nips/BrownS17}. The results in Table~\ref{table:t1} showcase that after conducting $600,000$ hands, RL-CFR achieves $64$ mbb/hand win-rate against ReBeL's replication.

    \begin{table}[!htp]

      \caption{Competition results of the HUNL AIs against each other, measured in mbb/hand (variance was reduced by AIVAT technique \cite{DBLP:conf/aaai/BurchSMMB18}).}
      \label{table:t1}
      \begin{center}
      \begin{tabular}{ccc}
        AI name & ReBeL (Replication) & Slumbot \\
        \hline\\
        ReBeL (Replication) & - & $18\pm 16$\\
        ReBeL \cite{DBLP:conf/nips/BrownBLG20}& - & $45\pm 5$ \\
        \textbf{RL-CFR (Ours)} & $64\pm 11$ & $84\pm 17$ \\
        
      \end{tabular}
        \end{center}
    \end{table}
    
    We further compare RL-CFR against the open source AI Slumbot \cite{2013Slumbot}, which was the winner of the 2018 ACPC and is the only HUNL AI we know that offers online competition testing. Since the opponent may select actions that deviate from the game tree, we perform nested subgame solving technique \cite{DBLP:conf/nips/BrownS17} mentioned in Appendix \ref{abs}. 
    
    We play RL-CFR for over $250,000$ hands against Slumbot, and the test results are shown in Table~\ref{table:t1}, which illustrate that the replication of ReBeL beat Slumbot with a win-rate of $18$ mbb/hand, while RL-CFR beat Slumbot with a win-rate of $84$ mbb/hand, and the win-rate of RL-CFR against Slumbot improved by $66$ mbb/hand relative to ReBeL's replication. The performance of ReBeL's replication is worse than the original ReBeL version on the HUNL benchmark. This is because the  number of training samples of ReBeL's replication being less than those of the original ReBeL implementation \cite{DBLP:conf/nips/BrownBLG20} due to the limited computational resources. We emphasize that ReBeL is a building block of our novel RL-CFR scheme, and our results show that RL-CFR achieves significant win-rate against ReBeL's replication and Slumbot just by optimizing the action abstraction.\footnote{A win-rate of over $50$ mbb/hand in poker is a commonly cited benchmark for what a professional poker player seeks to win from a weaker opponent \cite{DBLP:journals/cacm/BowlingBJT17}.} 

    We also performed an exploitability evaluation in over $10,000$ random river stage states.\footnote{We simulate RL-CFR versus ReBeL's replication until reaching the river, meaning that the two players choose their respective action abstractions, and the performance of the previously chosen action abstraction has no effect on the test results.} The exploitability of a strategy $\sigma$ and a player $p$ is calculated by $\text{expl}_p(\sigma)=u_p^{\sigma}-\mathop{\min}_{\sigma^{*}\in\Sigma_{-p}}u_p^{\langle\sigma_p,\sigma^{*}\rangle}$ \cite{DBLP:books/daglib/0016248}. The exploitability of RL-CFR is $17\pm 0$ mbb/hand, and the exploitability of ReBeL's replication is $20\pm 0$ mbb/hand. The results indicate that RL-CFR generates action abstractions that are also less likely to be exploited in the context of generating more win-rate.

    We conduct additional experiments for RL-CFR, and the results are presented in Table~\ref{table:t2}. In these experiments, RL-CFR is compared against the method of choosing an optimal action abstraction among multiple fixed action abstractions (MUL-ACTION). MUL-ACTION operates by selecting an action abstraction with the highest value for the root PBS $\beta_r$ among three action abstractions $\mathcal{AA}_{1}(\beta_r),\mathcal{AA}_{2}(\beta_r),\mathcal{AA}_{3}(\beta_r)$.\footnote{We set $\mathcal{AA}_{1}(\beta_r)=\{F,C,A,0.5\times \text{pot},1\times \text{pot},2\times \text{pot}\},\mathcal{AA}_{2}(\beta_r)=\{F,C,A,0.25\times \text{pot},0.5\times \text{pot},1\times \text{pot}\},\mathcal{AA}_{3}(\beta_r)=\{F,C,A,0.33\times \text{pot},0.7\times \text{pot},1.5\times \text{pot}\}$. Here the action abstractions other than the root are the same as those used in ReBeL.} RL-CFR outperforms MUL-ACTION by $21$ mbb/hand in win-rate after $100,000$ hands, requiring only 1/3 of the running time.

    Additionally, we compare RL-CFR against choosing a finer-grained action abstraction (FINE-GRAIN) $\mathcal {AA}(\beta_r)=\{F,C,A,0.25\times \text{pot},0.5\times \text{pot},0.75\times \text{pot},1.0\times \text{pot},1.25\times \text{pot},2.0\times \text{pot}\}$ at the root PBS $\beta_r$.\footnote{Following the same setting as in \cite{DBLP:journals/corr/abs-2007-10442}, and the action abstractions other than the root are the same as those used in ReBeL.} RL-CFR surpasses FINE-GRAIN by $23$ mbb/hand in win-rate after $100,000$ hands, requiring approximately $2/3$ of the running time.\footnote{Since the numbers of non-terminal nodes extended by the root node in the game tree built by FINE-GRAIN and RL-CFR are $9$ and $6$, respectively.} The results of RL-CFR versus fixed action abstraction methods illustrate that the action abstraction chosen by RL-CFR can effectively increase the win-rate without increasing the running time of the CFR algorithm.

    \begin{table}[!htp]
       
      \caption{
      Competition results of fixed action abstraction methods against RL-CFR.}
       \label{table:t2}
      \begin{center}
      \begin{tabular}{ccc}
        Method & Win-rate & Running time \\
        \hline\\
        ReBeL(Replication) & $-64\pm 11$ & $1\times$\\
        MUL-ACTION&  $-21\pm 26$& $3\times $ \\
        FINE-GRAIN & $-23\pm 28$ & $1.5\times$ \\
        
      \end{tabular}
        \end{center}
    \end{table}

\section{Conclusions}

    In this study, we introduce RL-CFR, a pioneering algorithmic solution designed for the resolution of large-scale IIEFGs. Anchored in a unique abstract MDP formulation, RL-CFR employs public information as states, leverages action abstraction features as actions, and adopts a meticulously crafted reward function. The ingenuity of RL-CFR lies in its fusion of reinforcement learning for action abstraction selection with CFR, facilitating dynamic action abstraction selection in IIEFGs. Our extensive experiments conducted in HUNL demonstrate that RL-CFR attains a substantial performance improvement when juxtaposed with fixed action abstraction HUNL methods. These findings underscore the efficacy and promise of RL-CFR in tackling the complexities inherent in large-scale Imperfect Information Extensive-Form Games.
    
\bibliographystyle{unsrt}  
\bibliography{RL_CFR}  
\newpage

\appendix

\startcontents[section]
\printcontents[section]{l}{1}{\setcounter{tocdepth}{2}}

\section{Texas Hold'em Rules}
\label{rule}

    The game begins with each player receiving two cards, referred to as their ``private hand." The gameplay unfolds through four stages: pre-flop, flop, turn, and river. During these stages, a total of five public cards are revealed – three at the start of the flop, one at the start of the turn, and one at the start of the river.

    Before the game commences, several players are required to contribute a pre-specified number of chips, known as the ``small blind" and ``big blind." Typically, the small blind is set at half the value of the big blind. In the pre-flop stage of HUNL, the small blind player takes the first action, while in other stages, the big blind player initiates the action. The permissible actions include fold, check/call, and bet/raise. In No-limit Texas Hold'em, players can bet or raise any number of chips between the last bet/raise in the stage (at least 1 big blind) and their remaining chips, even opting for an ``all-in."

    At the conclusion of a game, players who did not fold by the end of all stages select the best five cards from their private hand and the five public cards for comparison. The player, or players, with the best hand win the pot. The win-rate in Texas Hold'em is often expressed as the average number of big blinds won per game. Alternatively, it can be measured in more granular units such as mbb \cite{DBLP:journals/cacm/BowlingBJT17}, equivalent to one thousandth of a big blind. For instance, a win-rate of 0.01 big blind per hand can also be stated as 10 mbb/hand (10 mbb per hand).

\section{Related Work of Texas Hold'em AIs}
\label{ai}

In the realm of Texas Hold'em AI, substantial progress has been made, with notable achievements by powerful agents like Libratus \cite{DBLP:conf/ijcai/BrownS17} and Pluribus \cite{2019Superhuman}, demonstrating supremacy over top human players in both two-player and multi-player poker. These AI models employed intricate abstractions \cite{DBLP:conf/aaai/JohansonBBB12, DBLP:conf/aaai/GanzfriedS14, DBLP:conf/aaai/BrownGS15a} to handle the vast decision space of Texas Hold'em, requiring extensive computational resources for pre-calculating a blueprint strategy \cite{DBLP:conf/aaai/BrownS16} through CFR under an extensive game tree.

DeepStack \cite{DBLP:journals/corr/MoravcikSBLMBDW17} introduced deep learning to estimate the values of private hands within the game tree, effectively reducing its size  \cite{DBLP:journals/corr/abs-1302-7008}. ReBeL \cite{DBLP:conf/nips/BrownBLG20} leveraged self-play reinforcement learning \cite{DBLP:phd/ethos/Heinrich17} to generate realistic training data, providing an alternative approach to training. Notably, both Libratus and Pluribus did not incorporate reinforcement learning, while DeepStack and ReBeL did not employ reinforcement learning in action abstraction selection, specifically raising scales in HUNL. A recent contribution by \cite{DBLP:conf/aaai/ZhaoYLLX22} presented a HUNL AI based on reinforcement learning, showcasing the potential for creating excellent AIs with minimal computational resources. However, the absence of the widely used CFR algorithm in HUNL limits its theoretical guarantees and raises concerns about exploitability.

In summary, the landscape of Texas Hold'em AI is marked by diverse approaches, encompassing complex abstractions, CFR, deep learning, and reinforcement learning. Ongoing research aims to strike a balance between computational efficiency and strategic sophistication in order to further advance the capabilities of AI agents in this challenging poker variant.

\section{Counterfactual Regret Minimization}
    \label{cfr}

        Counterfactual Regret Minimization (CFR) is an algorithm tailored for large IIEFGs, aimed at minimizing regret independently within each information set \cite{DBLP:conf/nips/ZinkevichJBP07}. It is capable of finding $\varepsilon$-Nash equilibrium in two-player zero-sum IIEFGs.

        Let $\sigma^t$ represent the strategy profile at iteration $t$. The instantaneous regret for taking action $a$ at information set $I \in \mathcal I_p$ in iteration $t$ is denoted as $r^t(I, a) = v_p^{\sigma^t}(I, a) - v_p^{\sigma^t}(I)$. The counterfactual regret for choosing action $a$ at $I$ in iteration $T$ is defined as $R^T(I, a) = \sum{t=1}^T r^t(I, a)$. This counterfactual regret is used in regret matching (RM) \cite{1997A}, a no-regret learning algorithm employed for solving imperfect-information games.
        
        For an information set $I$, on each iteration $t+1$, an action $a\in\mathcal {AA}(I)$ is chosen based on probabilities $\sigma^{t+1}(I,a)=\frac{R^t_+(I,a)}{\sum_{a'\in \mathcal {AA}(I)}R^t_+(I,a')}$ where $R^t_+(I,a)=\mathop{\max}\{0,R^t(I,a)\}$. If $\sum_{a'\in\mathcal {AA}(I)}R^t_+(I,a')=0$, an arbitrary strategy can be chosen. Generally, the upper bound on regret values for CFR or its variants  \cite{DBLP:conf/nips/LanctotWZB09,DBLP:journals/corr/Tammelin14,DBLP:conf/icml/BrownLGS19,DBLP:conf/aaai/BrownS19} is $O(L\sqrt{|\mathcal{AA}(I)|}\sqrt{T})$, where $L$ is the payoff range, $|\mathcal {AA}(I)|$ is the size of action abstraction for information set $I$ and $T$ is the number of iterations \cite{DBLP:books/daglib/0016248}.
        
        Discounted CFR (DCFR) \cite{DBLP:conf/aaai/BrownS19} stands out as a prominent equilibrium-finding algorithm for large IIEFGs \cite{Brownthesis}. DCFR is a variant of CFR with parameters $\alpha,\beta,\gamma$ (DCFR$_{\alpha,\beta,\gamma}$). Specifically, accumulated positive regrets are multiplied by $\frac{t^{\alpha}}{t^{\alpha}+1}$, negative regrets by $\frac{t^{\beta}}{t^{\beta}+1}$, and contributions to the average strategy $\overline{\sigma}$ by $(\frac{t}{t+1})^{\gamma}$ on each iteration $t$. In our experiments, we set $\alpha=\frac{3}{2}$, $\beta=0$, and $\gamma=2$, denoted as $\text{DCFR}_{\frac{3}{2},0,2}$.

\section{Abstraction}  

        \label{abs}

        The huge solution complexity of IIEFGs is characterized by three dimensions: the depth of the game $D$, the size of the information set $|I|$ and the number of available actions $|\mathcal A(I)|$.The original space complexity is $O(|\mathcal A(I)|^D\cdot|I|)$, reaching over the order of $10^{160}$ for HUNL with stacks of $200$ big blinds and $20,000$ chips \cite{DBLP:journals/corr/abs-1302-7008}. The time complexity of CFR to solve an IIEFG is $O(T\cdot|\mathcal A(I)|^D\cdot|I|)$ where $T$ is the number of iterations.

        To limit the depth of the game, it is common practice not to compute the strategy until the end. Instead, a depth-limited subgame \cite{DBLP:conf/nips/BrownSA18} is generated, extending only a limited number of states into the future. Strategies or expected values are estimated for leaf states, which are non-terminal states in the full game but terminal states in the depth-limited subgame. DeepStack \cite{DBLP:journals/corr/MoravcikSBLMBDW17} and ReBeL \cite{DBLP:conf/nips/BrownBLG20} employ deep learning to estimate counterfactual values of leaf states, thereby avoiding solving until the end of the game. Another approach to limit depth involves consuming substantial computing resources to pre-calculate a blueprint strategy \cite{DBLP:conf/aaai/BrownS16, DBLP:conf/ijcai/BrownS17}, avoiding extensive solving for deep game instances.

        To limit the size of the information set, similar states can be grouped into the same bucket (information abstraction) \cite{DBLP:conf/aaai/JohansonBBB12, DBLP:conf/atal/JohansonBVB13, DBLP:conf/aaai/BrownGS15a} or represented in a high-dimensional feature abstraction \cite{DBLP:conf/icml/BrownLGS19}.\footnote{These methods also reduce the number of nodes in the game tree by grouping similar nodes into the same bucket.} Information abstractions require careful design tailored to the specific game. To demonstrate the generality of our method for general IIEFGs, our experiments refrain from using any state-space abstractions.

        To restrict the number of available actions, it is common to utilize action abstraction in IIEFGs. Formally, $\mathcal{AA}(I)$ represents the set of available actions at information set $I$, and $\mathcal{AA}(I) \subseteq \mathcal{A}(I)$ is an action abstraction for $\mathcal{A}(I)$. If the opponent chooses an off-tree action $a$ not in the action abstraction $\mathcal{AA}(I)$, rounding off-tree action to a nearby in-abstraction action \cite{DBLP:conf/ijcai/SchnizleinBS09, DBLP:conf/ijcai/GanzfriedS13} or resolving the strategy based on the new action abstraction $\mathcal{AA}(I) \cup {a}$ (nested subgame solving \cite{DBLP:conf/atal/GanzfriedS15, DBLP:conf/nips/BrownS17}) is commonly employed.

\section{Solving the strategy and PBS value for large IIEFGs}
\label{app1}
    In this section, we introduce the training process of the ReBeL algorithm \cite{DBLP:conf/nips/BrownBLG20}, a self-play RL method for solving the strategy and PBS values for large IIEFGs. We use HUNL as an example to describe the setting of specific parameters.

    In each epoch, training commences from the initial state of the game, and the PBS corresponding to the initial state is denoted as $\beta_{\text{init}}$. During training, we handle a PBS $\beta_r$\footnote{In general, PBS can shed extraneous history to refine information \cite{DBLP:conf/nips/BrownBLG20}.} and its corresponding action abstraction $\mathcal{AA}(\beta_r)$. The task involves computing the PBS value $v(\beta_r)$ and sampling to a leaf PBS $\beta_z$. Algorithm \ref{a2} details these steps, and we provide a description of the training process below.

    \begin{algorithm}[htb]
     
    \caption{ReBeL \cite{DBLP:conf/nips/BrownBLG20} algorithm: Solving the strategy and PBS value for PBS $\beta_r$ with action abstraction $\mathcal{AA}(\beta_r)$}
    \label{a2}

    \begin{algorithmic}
        \Function{ReBeL}{$\beta_r,\mathcal {AA}(\beta_r)$}
        
            \State $G\leftarrow$ConstructSubgame$(\beta_r,\mathcal {AA}(\beta_r))$\Comment{\textcolor{blue}{Construct a subgame rooted with $\beta_r$}}

            \State $\overline{\sigma},\sigma^0\leftarrow$UniformPolicy$(\beta_r,\mathcal {AA}(\beta_r))$

            \State $\mathbf{v}(\beta_r)\leftarrow \mathbf{0}$

            \State $t_{\text{sample}}\sim \text{unif}\{1,T\}$\Comment{\textcolor{blue}{Sample next iteration}}

            \For{$t=1\cdots T$}

                \State $G\leftarrow$LeafValueEstimate$(G,\sigma^{t-1},\theta)$\Comment{\textcolor{blue}{$\theta$ is the parameters of PBS value network}}

                \State $\sigma^t\leftarrow$UpdatePolicy$(G,\sigma^{t-1})$

                \State $\overline{\sigma}\leftarrow\frac{t-1}{t+1}\overline{\sigma}+\frac{2}{t+1}\sigma^t$\Comment{\textcolor{blue}{Update average strategy based on $DCFR_{\frac{3}{2},0,2}$}}

                \State $\mathbf{v}(\beta_r)\leftarrow \frac{t-1}{t+1}\mathbf{v}(\beta_r)+\frac{2}{t+1}\mathbf{v}^{\sigma^t}(\beta_r)$\Comment{\textcolor{blue}{Update PBS value for all infostates at $\beta_r$}}

                \If{$t=t_{\text{sample}}$}
                
                \State $\beta_{\text{next}}\leftarrow$SampleLeaf$(G,\sigma^t)$\Comment{\textcolor{blue}{Sample a leaf PBS}}

                \EndIf

            \EndFor

            \State Add $\{\beta_r,\mathbf{v}(\beta_r)\}$ to $Data^{\text{PBS}}$\Comment{\textcolor{blue}{Add PBS data for training}}

            \State $v_{\mathcal P(\beta_r)}^{\overline{\sigma}}(\beta_r)\leftarrow$ComputeValue$(\mathbf{v}(\beta_r))$\Comment{\textcolor{blue}{Compute PBS value for acting player at $\beta_r$}}
        
            \Return{$\overline{\sigma},v_{\mathcal P(\beta_r)}^{\overline{\sigma}}(\beta_r),\beta_{\text{next}}$}
                
        \EndFunction
    \end{algorithmic} 

    \end{algorithm}

    At the onset of training, we construct a depth-limited subgame rooted with $\beta_r$.\footnote{In the HUNL experiments, the subgame is built up to the end of the two players' actions in a stage or the end of the chance player's action. This implies that an epoch has up to $7$ phases: start of pre-flop, end of pre-flop, start of flop, end of flop, start of turn, end of turn, and start of river.} During the construction of the game tree, for non-terminal and non-leaf node $\beta'$, we expand the child nodes downwards according to the action abstraction $\mathcal{AA}(\beta')$.\footnote{In the HUNL experiments, to reduce the size of the game tree, for non-terminal PBS $\beta$ other than the root and root's sons, we set $\mathcal{AA}(\beta)=\{F,C,A,0.8\times \text{pot}\}$.}

    Once the game tree is built, the subgame is solved by running $T$ iterations of the CFR algorithm. The value of leaf nodes is estimated using a learned value network $\hat{v}$ at each iteration based on their PBS. On each iteration $t$, CFR is employed to determine a strategy profile $\sigma^t$ in the subgame. Subsequently, the infostate value of a leaf node $z$ is set to $\hat{v}(O_p(z)|\beta_z^{\sigma^t})$, where $\beta_z^{\sigma^t}$ is the PBS at $z$ when players play according to $\sigma^t$.

    Due to the non-zero-sum nature of estimates from the neural network, adjustments are made to the infostate values at each PBS to ensure the game satisfies the zero-sum property. Additionally, for infostates that should have the same value, their value estimates are averaged. Since PBSs may change every iteration, the leaf node values may also change. Given $\sigma^t$ and leaf node values, each infostate in each node has a calculated PBS value, as explained in Section \ref{back}. This information is then used to update the regret and average strategy $\overline{\sigma}$ for the CFR algorithm.

    After completing $T$ iterations, we obtain the solved average strategy $\overline{\sigma}$. Using this strategy, we calculate the PBS values for all infostates $v_p^{\overline{\sigma}}(O_p|\beta_r)$ for the root PBS $\beta_r$. We denote this vector of PBS values as $\mathbf{v}(\beta_r)$. Subsequently, we add the PBS data ${\beta_r,\mathbf{v}(\beta_r)}$ to the training data (denoted $Data^{\text{PBS}}$) for $\hat{v}(\beta_r)$. Meanwhile, we calculate the PBS value $v_{\mathcal{P}(\beta_r)}^{\overline{\sigma}}$ of $\beta_r$ according to the calculated value vector $\mathbf{v}(\beta_r)$. This PBS value is required as part of the reward function for our RL-CFR framework, although it is not needed in the ReBeL algorithm itself.

    Next, we sample a leaf PBS $\beta_z$ according to $\sigma^t$ on a random iteration $t \sim \text{unif}\{1,T\}$, where $T$ is the number of iterations. To ensure more exploration, we can sample a random leaf PBS with probability $\varepsilon$. Additionally, we can modify some public information in the sampled PBS for more exploration.\footnote{For HUNL agent training, we set $\varepsilon=25\%$. For a sampled PBS, we multiply the chips in the pot by a random number between $0.9$ and $1.1$. For the PBS corresponding to the initial state, we set the chips of all players to a random number between $50$ big blinds and $250$ big blinds.} We repeat the above processes until the game ends.
    

    \begin{figure}[!htb]
  \centering
  
  \includegraphics[width=\textwidth,trim=100 270 100 30]{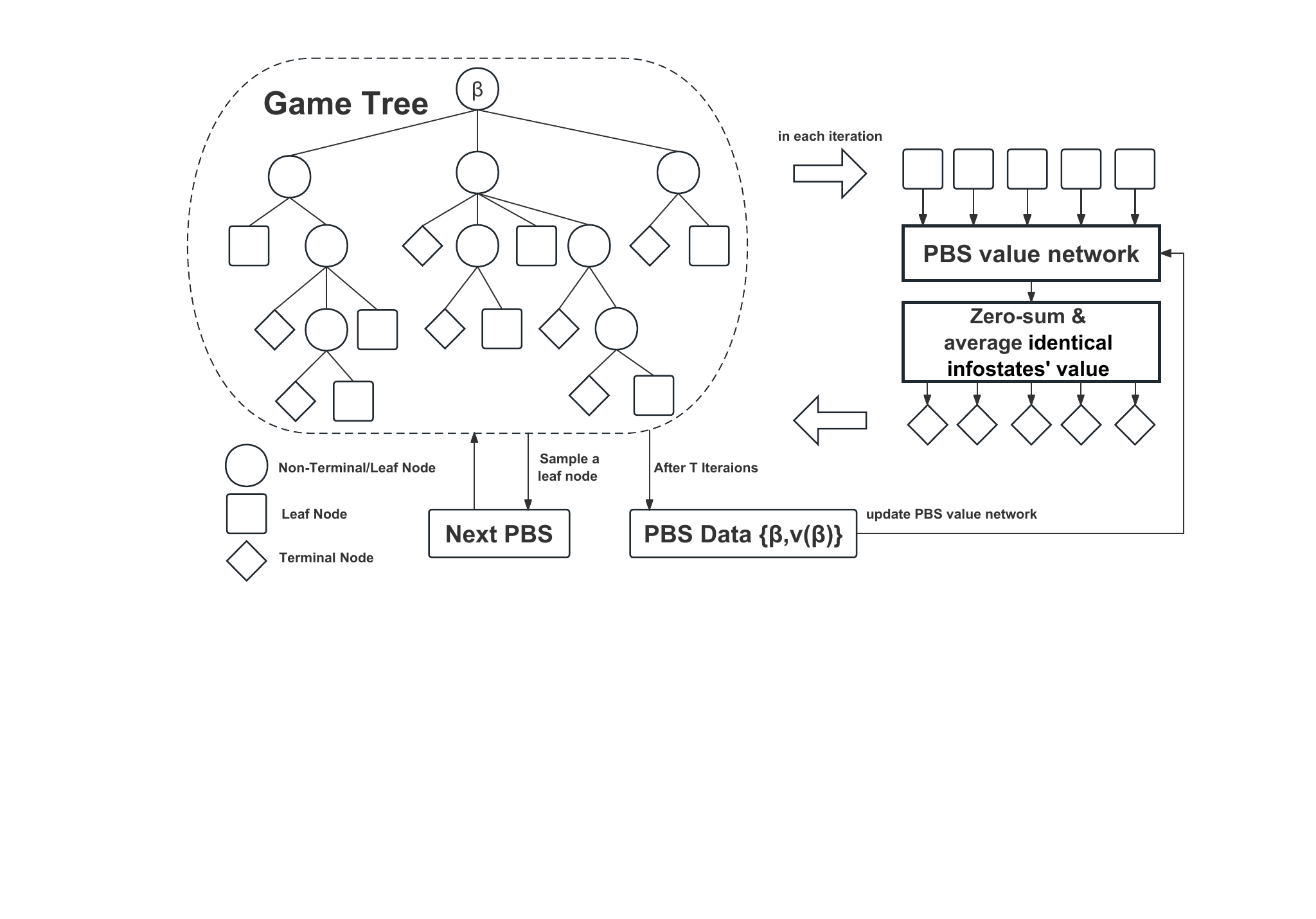}
  \caption{This figure illustrates the process of generating PBS data and training the PBS value network. For a given PBS $\beta$, we construct a depth-limited subgame rooted with $\beta$. Non-terminal and non-leaf nodes are depicted as circles, and during the construction of the game tree, we expand child nodes based on the action abstraction of the PBS associated with the node. Terminal nodes, denoted by diamonds, allow for direct calculation of the PBS value. Leaf nodes, represented by rectangles, require the estimation of PBS values in every iteration of CFR, where the PBS value network is employed to estimate the PBS values for these leaf nodes.
  }
  
  \label{p2:2}
  \end{figure}

    We utilize Huber Loss \cite{10.1214/aoms/1177703732} as the loss function for the PBS value network: 
    \begin{equation}
    \begin{aligned}
        \mathcal L(\theta,\delta)=
        &
        \mathbb E_{\{O_p,v_p(O_p)\}\sim\{\beta_r,\mathbf{v}(\beta_r)\},\{\beta_r,\mathbf{v}(\beta_r)\}\sim Data^{\text{PBS}}}\\
        &[\mathop{\min}\{\frac{1}{2}(v_p(O_p)-\hat{v}^{\theta}(O_p|\beta_r))^2,\delta|v_p(O_p)-\hat{v}^{\theta}(O_p|\beta_r)|-\frac{1}{2}\delta^2\}]
    \end{aligned}
    \end{equation}

    where $\theta$ represents the parameters of the PBS value network, $O_p$ is an infostate in PBS $\beta_r$, and $\delta$ is a hyperparameter of the Huber Loss.

    In our HUNL experiments, the PBS value network consists of $6$ layers and $18$ million parameters. The input layer has $2,678$ dimensions, corresponding to all possible private hands of the two players and the public state information. Each hidden layer comprises $1,536$ dimensions, and the output layer has $2,652$ dimensions, representing all possible private hands of the two players. We sampled $6\times 10^7$ PBS data during training. Random sampling was performed from the last $1\times 10^7$ PBS data, and the training employed a learning rate of $1\times 10^{-5}$ with a batch size of $512$. The training process and data sampling occurred simultaneously. Following the training of PBS value networks, we obtained a replication version of ReBeL as a baseline.

    In summary, ReBeL is a self-play RL framework capable of continuously generating data from scratch for training. Figure \ref{p2:2} illustrates the training process of the ReBeL algorithm. 
    
    \section{Examples of RL-CFR strategies}
    
    We use several examples from HUNL to illustrate how RL-CFR selects action abstractions. We show examples of heads-up evaluation between ReBeL's replication and RL-CFR. Both players start with $200$ big blinds (BB) and $20,000$ chips ($100$ chips for $1$ BB) in all examples.

    \textbf{Example 1.}
    
    \emph{Pre-flop stage.} ReBeL sits in small blind position with hand $4\spadesuit3\spadesuit$ and RL-CFR sits in big blind position with hand $J\heartsuit8\diamondsuit$. ReBeL acts first with action abstraction $\{F,C,2,3,5,A\}$\footnote{$F,C,A$ refer to fold, check/call and all-in respectively and the numbers represent raising scales in BB.}. The strategy calculated by CFR is: call with $3.21\%$, raise to $2$ BB ($0.5\times$pot) with $52.10\%$, raise to $3$ BB ($1\times$pot) with $44.11\%$ and raise to $4$ BB ($2\times$pot) with $0.58\%$. ReBeL raises to $2$ BB in this example. In this situation, RL-CFR selects an action abstraction $\{F,C,3,8.8,16.21,A\}$. The strategy is: call with $76.08\%$, raise to $8.8$ BB ($1.7\times$pot) with $23.67\%$ and raise to $16.21$ BB ($3.5525\times$pot) with $0.25\%$. ReBeL calls in this example. When RL-CFR in the big blind is faced with a $2$ BB raise, RL-CFR will use the three raising scales of $3$ BB, $8.8$ BB, $16.21$ BB and expect to win $10$ mbb/hand compared to the default raising scales ($4$ BB, $6$ BB, $10$ BB).

    \emph{Flop stage.} Flop is $J\diamondsuit 6\heartsuit 3\diamondsuit$. There are $4$ BB in the pot and RL-CFR acts first. RL-CFR selects an action abstraction $\{F,C,A\}$. In this case, RL-CFR will check all hands, which is a common strategy that human professional players will employ. Now turn to ReBeL and the strategy is: check with $47.94\%$, bet $2$ BB ($0.5\times$pot) with $51.63\%$ and bet $4$ BB ($1\times$pot) with $0.42\%$. ReBeL bets $2$ BB in this example. In this situation, RL-CFR selects an action abstraction $\{F,C,4,25.36,A\}$. The strategy is: call with $72.09\%$ and raise to $4$ BB ($0.25\times$pot) with $27.91\%$. RL-CFR calls in this example. It's an interesting strategy, with RL-CFR opting for a minimum raising scale (mini-raise) and a very large raising scale, gaining an additional $6$ mbb/hand win-rate compared to the default action abstraction.

    \emph{Turn stage.} Turn is $4\clubsuit$. There are $8$ BB in the pot and RL-CFR acts first. RL-CFR selects an action abstraction $\{F,C,1,A\}$. The strategy is: check with $20.64\%$ and bet $1$ BB with $79.36\%$. The turn card is very favourable to RL-CFR's calling range, so RL-CFR had a high frequency of betting (donk). Choosing a $1$ BB raising scale (minimum betting) gives RL-CFR an additional win-rate of $42$ mbb/hand compared to the default action abstraction, which is very impressive. RL-CFR bets $1$ BB in this example. Now turn to ReBeL and the strategy is: call with $99.63\%$ and raise to $6$ BB ($0.5\times$pot) with $0.37\%$. ReBeL has two-pairs now, however the strategy calculated by CFR is calling most hands since the turn card is unfavorable to small blind player's hand range.

    \emph{River stage.} River is $2\heartsuit$. There are $10$ BB in the pot and RL-CFR acts first. RL-CFR selects an action abstraction $\{F,C,8.42,14.88,46.22,A\}$ with $4$ mbb/hand extra win-rate. The strategy is: check with $99.93\%$ and bet $8.42$ BB ($0.842\times$pot) with $0.07\%$. Now turn to ReBeL and the strategy is: check with $0.37\%$, bet $5$ BB ($0.5\times$pot) with $43.76\%$, bet $10$ BB ($1\times$pot) with $55.66\%$ and bet $20$ BB ($2\times$pot) with $0.19\%$. ReBeL bets $5$ BB in the example. Now turn to RL-CFR and the selected action abstraction is $\{F,C,17.81,55.31,98.77,A\}$  with $6$ mbb/hand extra win-rate. The strategy of RL-CFR is: fold with $42.04\%$, call with $56.52\%$, raise to $17.81$ BB ($0.6405\times$pot) with $0.55\%$ and raise to $93.77$ BB ($4.4385\times$pot) with $0.87\%$. RL-CFR folds and loses the $10$ BB pot in the example.

    \textbf{Example 2.} It is a symmetrical example of Example 1.

    \emph{Pre-flop stage.} RL-CFR sits in small blind position with hand $4\spadesuit3\spadesuit$ and ReBeL sits in big blind position with hand $J\heartsuit8\diamondsuit$. RL-CFR acts first and selects an action abstraction $\{F,C,2.9,4.56,7.64,A\}$ with $24$ mbb/hand extra win-rate. The strategy of RL-CFR is: call with $18.16\%$, raise to $2.9$ BB ($0.95\times$pot) with $78.54\%$ and raise to $4.56$ BB ($1.78\times$pot) with $3.29\%$. RL-CFR raises to $2.9$ BB and ReBeL calls in this example.

    \emph{Flop stage.} Flop is $J\diamondsuit 6\heartsuit 3\diamondsuit$. There are $5.8$ BB in the pot and ReBeL checks first. The action abstraction selected by RL-CFR is $\{F,C,8.6,28.1,A\}$. It is an interesting strategy generated by RL-CFR with overbet (bet more than a pot) only and gain an additional $46$ mbb/hand win-rate compared to the default action abstraction. The strategy is: check with $99.65\%$, bet $8.6$ BB with $0.22\%$ and bet $28.1$ BB with $0.13\%$. RL-CFR checks in the example.

    \emph{Turn stage.} Turn is $4\clubsuit$. There are $5.8$ BB in the pot and ReBeL acts first. The strategy of ReBeL is: check with $60.17\%$, bet $2.9$ BB with $26.06\%$, bet $5.8$ BB with $13.66\%$ and bet $11.6$ BB with $0.12\%$. ReBeL checks in the example and turn to RL-CFR. The action abstraction calculated by RL-CFR is $\{F,C,1,1.85,A\}$. However, after evaluation by the policy network, RL-CFR considers this action abstraction to be inferior to the default action abstraction, so the default action abstraction will be chosen this time\footnote{After selecting an action abstraction through the action network, we evaluate it using the policy network and if the evaluation value is negative, we use the default action abstraction.}. The strategy of RL-CFR is: check with $0.03\%$, bet $2.9$ BB ($0.5\times$pot) with $98.88\%$, bet $5.8$ BB ($1\times$pot) with $0.14\%$ and bet $11.6$ BB ($2\times$pot) with $0.94\%$. In this example, RL-CFR bets $2.9$ BB and ReBeL calls.

    \emph{River stage.} River is $2\heartsuit$. There are $11.6$ BB in the pot and ReBeL checks first. The action abstraction selected by RL-CFR is $\{F,C,2.3,44.41,46.43\}$ and the strategy of RL-CFR is: check with $0.30\%$, bet $2.3$ BB with $99.64\%$ and bet $44.41,44.63$ BB with $0.06\%$. In this example, RL-CFR bets $2.3$ BB and ReBeL calls. RL-CFR wins the $16.2$ BB pot with two pairs at showdown.

    \textbf{Example 3.} In this example RL-CFR performed a bluff (betting with a weaker hand) and successfully bluffing with a suitable action abstraction.

    \emph{Pre-flop stage.} ReBeL sits in small blind position with hand $Q\heartsuit9\heartsuit$ and RL-CFR sits in big blind position with hand $9\clubsuit8\heartsuit$. ReBeL raises $2$ BB first and RL-CFR calls in this example.

    \emph{Flop stage.} Flop is $K\clubsuit6\spadesuit2\diamondsuit$. There are $4$ BB in the pot. RL-CFR and ReBeL check in this example.

    \emph{Turn stage.} Turn is $7\clubsuit$. There are $4$ BB in the pot and RL-CFR acts first. The action abstraction selected by RL-CFR is $\{F,C,1,2.25,A\}$ with $10$ mbb/hand extra win-rate. The strategy of RL-CFR is: check with $34.31\%$, bet $1$ BB with $11.06\%$ and bet $2.25$ BB with $54.63\%$. RL-CFR bets $1$ BB in this example. The stragety of ReBeL is: fold with $48.32\%$, call with $51.04\%$, raises to $4$ BB with $0.61\%$ and raises to $7$ BB with $0.03\%$. ReBeL calls in this example.

    \emph{River stage.} River is $4\spadesuit$. There are $6$ BB in the pot and RL-CFR acts first. The action abstraction selected by RL-CFR is $\{F,C,4.56,5.71,29.23,A\}$ with $3$ mbb/hand extra win-rate. The strategy of RL-CFR is: check with $12.48\%$, bet $4.56$ BB with $41.63\%$, bet $5.71$ BB with $33.11\%$ and bet $29.23$ BB with $12.78\%$. RL-CFR bets $5.71$ BB in this example and the strategy of ReBeL is: fold with $99.14\%$ and call with $0.85\%$. ReBeL folds and RL-CFR wins the $6$ BB pot.

    \textbf{Example 4.} In this example RL-CFR calls the 3-bet (re-raise at pre-flop) from ReBeL.

    \emph{Pre-flop stage.} RL-CFR sits in small blind position with hand $A\diamondsuit4\diamondsuit$ and ReBeL sits in big blind position with hand $K\spadesuit J\spadesuit$. RL-CFR raises to $2.9$ BB. The strategy of ReBeL is: call with $2.37\%$, raises to $5.8$ BB with $36.31\%$, raises to $8.7$ BB with $42.01\%$ and raises to $14.5$ BB with $19.31\%$. ReBeL raises to $8.7$ BB in this example. The action abstraction selected by RL-CFR is $\{F,C,14.5,16.22,27.96,A\}$ with $13$ mbb/hand extra win-rate. The strategy of RL-CFR is: call with $70.90\%$, raise to $14.5$ BB with $0.03\%$, raise to $16.22$ BB with $0.04\%$, and raise to $27.96$ BB with $29.03\%$. RL-CFR calls in the example.

    \emph{Flop stage.} Flop is $Q\clubsuit 5\diamondsuit 3\diamondsuit$. There are $17.4$ BB in the pot and ReBeL acts first. The strategy of ReBeL is: check with $80.33\%$, bet $8.7$ BB with $14.94\%$, bet $17.4$ BB with $4.46\%$ and bet $34.8$ BB with $0.26\%$. ReBeL bets $17.4$ BB in the example. The action abstraction selected by RL-CFR is $\{F,C,34.8,A\}$ with $142$ mbb/hand extra win-rate, which means that if there is no mini-raise in the action abstraction there will be a huge loss in this situation. The strategy of RL-CFR is: call with $97.69\%$, raise to $34.8$ BB with $1.73\%$ and all-in with $0.58\%$. RL-CFR calls in the example.

    \emph{Turn stage.} Turn is $6\diamondsuit$ and RL-CFR has the nuts (strongest hand). There are $52.2$ BB in the pot and ReBeL checks first. The action abstraction selected by RL-CFR is $\{F,C,15.11,64.73,97.22,A\}$ with $46$ mbb/hand extra win-rate. The strategy of RL-CFR is: check with $0.43\%$, bet $15.11$ BB with $88.13\%$ and bet $64.73$ BB with $11.43\%$. RL-CFR bets $15.11$ BB in the example and ReBeL folds. RL-CFR wins the $52.2$ BB pot. The action abstraction of RL-CFR in this situation is very reasonable, and a $15.11$ BB bet can put many of opponent's hands in an embarrassing situation.

\end{document}